\documentstyle[11pt]{article}

\renewcommand{\baselinestretch}{1.4}
\begin{document}

\def\slash{{\rlap /}}
\def\PRL{Phys. Rev. Lett. }
\def\be{\begin{equation}} \def\bea{\begin{eqnarray}}
\def\ee{\end{equation}}\def\eea{\end{eqnarray}}
\def\ncr{\nonumber\\ }
\def\tr{{\rm tr}\,}\def\STr{{\rm STr}\,}
\def\Sdet{{\rm Sdet}\,}\def\Tr{{\rm Tr}\,}
\def\dos{\stackrel{\star}{,}}
\def\DG{\Delta ^{\a\b\g}_{\s\r\m}\,}
\def\DR{\Delta ^{abc}_{srm}\,}
\def\nn{{\Box ^{-1}N_0}} \def\nj{{\Box ^{-1}N_1}}
\def\tj{{\Box ^{-1}T_1}} \def\td{{\Box ^{-1}T_2}}
\def\cpe{{i\over (4\pi )^2\epsilon} }
\def\jcpe{{1\over (4\pi )^2\epsilon} }

\def\MBVR{Maja Buri\'c\footnote{E-mail: majab@ff.bg.ac.yu} and
 Voja Radovanovi\'c\footnote{E-mail: rvoja@ff.bg.ac.yu}\\
{\it Faculty of Physics, University of Belgrade, P.O. Box 368,
11001 Belgrade, Serbia and Montenegro}}

\def\endtitle{\par\end{quotation}\vskip3.5in minus2.3in\newpage}


\def\a{\alpha}
\def\b{\beta}
\def\c{\chi}
\def\d{\delta}
\def\e{\epsilon}                
\def\f{\phi}                    
\def\g{\gamma}
\def\h{\eta}
\def\i{\iota}
\def\j{\psi}
\def\k{\kappa}
\def\l{\lambda}
\def\m{\mu}
\def\n{\nu}
\def\o{\omega}
\def\p{\pi}                     
\def\q{\theta}                  
\def\r{\rho}                    
\def\s{\sigma}                  
\def\t{\tau}
\def\u{\upsilon}
\def\x{\xi}
\def\z{\zeta}
\def\D{\Delta}
\def\F{\Phi}
\def\G{\Gamma}
\def\J{\Psi}
\def\L{\Lambda}
\def\O{\Omega}
\def\P{\Pi}
\def\Q{\Theta}
\def\S{\Sigma}
\def\U{\Upsilon}
\def\X{\Xi}
\def\pa{\partial}
\def\de{\nabla}

\def\ca{{\cal A}}
\def\cb{{\cal B}}
\def\cc{{\cal C}}
\def\cd{{\cal D}}
\def\ce{{\cal E}}
\def\cf{{\cal F}}
\def\cg{{\cal G}}
\def\ch{{\cal H}}
\def\ci{{\cal I}}
\def\cj{{\cal J}}
\def\ck{{\cal K}}
\def\cl{{\cal L}}
\def\cm{{\cal M}}
\def\cn{{\cal N}}
\def\co{{\cal O}}
\def\cp{{\cal P}}
\def\cq{{\cal Q}}
\def\car{{\cal R}}
\def\cs{{\cal S}}
\def\ct{{\cal T}}
\def\cu{{\cal U}}
\def\cv{{\cal V}}
\def\cw{{\cal W}}
\def\cx{{\cal X}}
\def\cy{{\cal Y}}
\def\cz{{\cal Z}}


\def\CMP{Commun. Math. Phys.}
\def\NP{Nucl. Phys. B\,}
\def\PL{Phys. Lett. B\,}
\def\PR{Phys. Rev. Lett.}
\def\PRD{Phys. Rev. D\,}
\def\CQG{Class. Quant. Grav.}
\def\IJMP{Int. J. Mod. Phys.}
\def\MPL{Mod. Phys. Lett.}



\topmargin=.17in                        
\headheight=0in                         
\headsep=0in                    
\textheight=9in                         
\footheight=3ex                         
\footskip=4ex           
\textwidth=6in                          
\hsize=6in                              
\parindent=21pt                         
\parskip=\medskipamount                 
\lineskip=0pt                           
\abovedisplayskip=1em plus.3em minus.5em        
\belowdisplayskip=1em plus.3em minus.5em        
\abovedisplayshortskip=.5em plus.2em minus.4em  
\belowdisplayshortskip=.5em plus.2em minus.4em  
\def\baselinestretch{1.2}       
\thicklines                         
\oddsidemargin=.25in \evensidemargin=.25in      
\marginparwidth=.85in                           


\def\title#1#2#3#4{
        {\hbox to\hsize{#4 \hfill  #3}}\par
        \begin{center}\vskip.5in minus.1in {\Large\bf #1}\\[.5in minus.2in]{#2}
        \vskip1.4in minus1.2in {\bf ABSTRACT}\\[.1in]\end{center}

     \begin{quotation}\par}
\def\author#1#2{#1\\[.1in]{\it #2}\\[.1in]}

\def\endabstract{\par\end{quotation}
        \renewcommand{\baselinestretch}{1.2}\small\normalsize}

\title{On renormalizability of the quantum electrodynamics on noncommutative space }
{\MBVR}{}{}

\noindent{In this paper we calculate the divergent part of the one
loop effective action for QED on noncommutative space using the
background field method. The effective action is obtained up to
the second order in the noncommutativity parameter $\theta$ and in
the classical fields.}

\endtitle

\section{Introduction}

The discovery of noncommutative (NC) structures in string theory
intensified the investigation of noncommutative field theories and
noncommutative geometry in the last couple of years. An important
aspect of this investigation is a representation-free
 formulation of the theory
 - endowing the abstract algebra of
noncommuting coordinates with structures like derivations, forms,
fields etc. \cite{mssw, jmssw, mad}. Various types of
noncommutativity can be analyzed in this way. The most often
considered one is the canonical structure, defined by
 \be\left[ \hat x^\mu , \hat x^\nu \right]=i\theta ^{\m\n}
 \ ,\label{can}\ee where
$\hat x^\m$, $\hat x^\n$ are the elements of the algebra, and
 $\theta^{\m\n}=-\theta^{\n\m}$ are constants of dimension
length$\, ^2$ .
 Other structures like Lie algebra or quantum plane
  have been discussed as well \cite{mssw, jmssw}. The other line
of work is to represent  NC theory by the
 fields on commutative
 space, encoding noncommutativity  in
 the definition of the product. The multiplication which
 corresponds to the canonical structure (\ref{can})
is the so-called Moyal-Weyl or $\star$-product: \be\label{moyal}
f\star g=e^{{i\over 2}\theta^{\m\n}{\pa\over \pa x^\m}{\pa\over
\pa y^\m}}f(x)g(y)|_{y\to x}\ ,\ee where $f$ and $g$ are functions
of the coordinates $x^\m$ on ${\bf R^4}$. Obviously,
\be\label{can1} [x^\m \dos x^\n]=x^\m\star x^\n-x^\n\star x^\m =
i\theta^{\m\n} \ .\ee It is possible to define $*$-products which
correspond to the other types of noncommutativity, too.

The   $\star$-product (\ref{moyal}) with its properties in
integration
  provides us with a new class of
 actions characterized by dimensionfull parameter $\theta$ and
 nonlocal lagrangians. In this setting the definition of
  noncommutative
 scalar field theories like $\Phi ^3$ or $\Phi ^4$ is
 straightforward. However, if one wants to define
a gauge theory, the use of noncommutative multiplication rule
imposes severe restrictions both on the choice of gauge group and
on the choice of representation \cite{h,bsst,cpst}. For example,
in NC electrodynamics  the values of  charge are quantized and
restricted to $\pm 1, 0$.

The result of Seiberg and Witten (SW) \cite{sw} on the equivalence
of classes of commutative and noncommutative gauge theories shows
that noncommutativity  is not equivalent to  quantization.
Noncommutative field theories can be quantized  in the
conventional perturbative way \cite{kw, mrs, mst}. By now, a
number of properties of NC field theories with constant $\theta$
are established. Some novel features appear, e.g. UV and IR
sectors are mixed in the perturbative expansion. The UV/IR mixing
can  be seen in the fact that 'nonplanar' diagrams contain terms
proportional to $\vert\tilde p\vert ^{-n}$, with $\tilde p^\mu
=\theta ^{\mu\nu}p_\nu$. In principle, one might expect that the
apparent nonrenormalizability of NC theories (seen already in the
fact that the 'coupling constant' $\theta $ is dimensionfull)
disappears after the summation of  perturbation series due to some
special properties of the lagrangian. This was indeed shown for
the NC $\Phi ^4$ theory in \cite{gp}; however, further
investigations seem to question this result \cite{GW} and prove
that the only renormalizable noncommutative theories are the
supersymmetric ones.

An obvious drawback of the perturbative treatment of NC theories
 is that the results, expressed as $\vert\tilde p\vert ^{-n}$,
are nonperturbative in the  parameter $\theta$. This means that
one cannot make a smooth commutative limit or estimate the effects
of noncommutativity in the lowest order (in the sectors where they
are small). In order to deal with this problem one uses  the
complementary approach of \cite{mssw} defined for
 the field theories with  gauge symmetry. As
 it is shown in \cite{mssw}, the representation of the gauge
 symmetry on NC space induces the expansion of noncommutative
 fields in the parameter $\theta$
in terms of fields on commutative space which coincides with the
SW expansion. The corresponding commutative 'physical' fields
carry the usual representation of  gauge symmetry on ${\bf R^4}$.
The SW map induces $\theta$-expansion in the action as well.
 This gives the possibility to treat $\theta$-linear
 and $\theta$-quadratic terms in the action as the first
corrections, describing the effects of noncommutativity in the
lowest order. Further, $\theta$-expanded action enables to
approach the problem of quantization in a different way, i.e.
considering the lagrangian order by order in $\theta$. In this
context, nonuniqueness of the SW map \cite{ak}  takes the role of
an additional 'symmetry'. Indeed, using the particular properties
of SW-expanded action, Bichl et al. \cite{bggpsw} succeeded to
prove full renormalizability of the  photon propagator for the
pure U(1) NC gauge theory. However, inclusion of the matter spoils
renormalizability \cite{bggpsw, bgpsw1, bgpsw2}, in spite of the
fact that diagrams in $\theta$-linear order have a high degree of
symmetry. In the case of massless fermions only one term breaks
renormalizability \cite{w}. This behavior motivates the
investigation of the theory in the $\theta ^2$-order.

Even if not renormalizable, $\theta$-expanded gauge theories can
be treated as effective theories at 'low' energies. Various
physical models, like noncommutative generalizations of the
standard model, have been proposed so far \cite{cjsww, AJSW}. In
that context also it is of importance to calculate the divergent
counterterms, as they describe the possible effective interaction
vertices.

This lecture is organized as follows. Since a large part of the
results on the $\theta ^2$-one-loop corrections of the propagators
were presented in \cite{BR}, we summarize them without entering
into details. We show how the additional contributions from
$\theta ^2$-classical lagrangian are obtained and thus complete
the result in the $\theta ^2$-order. We also give a somewhat more
extensive discussion of the field redefinitions and their
implications to renormalizability.

\section{Classical theory}

The noncommutative space which we use is  ${\bf R^4}$ with the
canonical structure \be\ [x^\m \dos x^\n]= i\theta^{\m\n} \ ,\ee
where $\m ,\n =0,\dots 3$
 and
$\star$ is the Moyal-Weyl product  (\ref{moyal}). The classical
action for electrodynamics  on this space is given by \be
\label{Snc}S=\int d^4x\, \hat{\bar \psi}\star (i\g^\m\hat
D_\m-m)\hat\psi -{1\over 4}\int d^4x \hat F_{\m\n}\star\hat
F^{\m\n}\ .\ee Here, $\hat\psi$ is the noncommutative fermionic
matter field while
 $\hat A_\m $ is the gauge potential. The corresponding
 field strength  $\hat F_{\m\n}$  is defined as \be \hat
F_{\m\n}=\pa_\m\hat A_\n-\pa_\n\hat A_\m-i(\hat A_\m\star\hat
A_\n-\hat A_\n\star\hat A_\m)\ ,\ee and the covariant derivative
$\hat D_\m\hat\psi$ \be \hat D_\m\hat\psi = \pa_\m\hat\psi -i\hat
A_\m\star \hat \psi\ .\ee

The fields $\hat\psi $, $\hat A_\m$, $\hat F_{\m\n}$ which give a
representation of noncommutative electrodynamics can be, via the
SW map,  mapped
 into the representation of ordinary U(1). To
the first orders in $\theta$ the map is given by \cite{jmssw}:
  \bea\label{sw} \hat
  A_\m&=&A_\m^{(0)}+A_\m^{(1)}+A_\m^{(2)}+\dots\\
  A_\m^{(0)}&=& A_\m\ncr
  A_\m^{(1)}&=& -{1\over 2}\,\theta^{\r\s}A_\r \left[(\pa_\s A_\m )+F_{\s\m}\right]\ncr
A_\m^{(2)}&=&{1\over 2}\,\theta^{\a\b}\theta^{\r\s}\left[ A_\a
A_\r(\pa _\s F_{\b\m})- (\pa _\b A_\m )(\pa _\r A_\a )A_\s +A_\a
F_{\b\r }F_{\s\m} \right]\ncr
 \label{spinor} \hat\psi&=&\psi ^{(0)}+\psi ^{(1)}+\psi
^{(2)}+\dots\\
  \psi ^{(0)}&=& \psi \ncr
  \psi ^{(1)}&=& -{1\over 2}\,\theta^{\r\s}A_\r(\pa_\s \psi )\ncr
\psi ^{(2)}&=&{1\over 32}\,\theta ^{\a\b }\theta^{\r\s }\big[
-4i(\pa _\a A_\r )(\pa _\b\pa _\s \psi )+4A_ \a A_\r (\pa _\b \pa
_\s \psi )+8A_\a (\pa _\b A_\r )(\pa _\s \psi )\ncr &+& 4A_\a
F_{\b\r } (\pa _\s \psi )-2(\pa _\a A_\r)(\pa _\b A_\s )\psi
+8iA_\a A_\s(\pa _\r A_\b )\psi \big]\ncr
 \hat
 F_{\m\n}&=&F_{\m\n}^{(0)}+F_{\m\n}^{(1)}+F_{\m\n}^{(2)}+\dots\\
F^{(0)}_{\m\n}&=& F_{\m\n}=\pa _\m A_\n -\pa _\n A_\m \ncr
F^{(1)}_{\m\n}&=&\theta ^{\r\s}\left[ F_{\m\r}F_{\n\s}-A_\r(\pa
_\s F_{\m\n})\right]\, ,\ {\rm etc.} \nonumber \eea
  Inserting (\ref{sw}) and (\ref{spinor}) into the action (\ref{Snc}),
  we get  the classical $\theta$-expanded action
 \cite{jmssw, bggpsw}
\be S =S^{(0)}+S _{A}^{(1)}+S_{\psi}^{(1)}+S
_{A}^{(2)}+S_{\psi}^{(2)} +\dots \ ,\ee with
 \bea\label{l0} S^{(0)}& =&\int
d^4x\,\Big[\bar\psi \big( i\g ^\m D_\m-m\big)\psi -{1\over
4}F^{\m\n}F_{\m\n}\Big]\ ,\\
 \label{lA} S _{A}^{(1)}&=&-{1\over
2}\,\theta ^{\r\s}\int d^4 x\,\Big[
F_{\m\r}F_{\n\s}F^{\m\n}-{1\over 4}F_{\r\s}F_{\m\n}F^{\m\n}\Big]\
,\\
  \label{l1psi} S_{\psi}^{(1)}&=&{1\over 2}\,\theta
^{\r\s}\int d^4x\,\Big[ -i F_{\m\r}\bar\psi\g ^\m D_\s\psi
+{1\over 2}F_{\r\s}\bar\psi (-i\g ^\m D_\m+m)\psi \Big] \ ,\\
 \label{l2A} S_{A}^{(2)}
&=&\int d^4 x\,\Big[ -{1\over 4}F^{(1)}_{\a\b}F^{(1)\a\b}-{1\over
2}F^{(0)}_{\a\b}F^{(2)\a\b}\Big]\\
 &=&\int d^4 x\,{1\over
4}\,\theta ^{\a\b}\theta ^{\m\n}\Big[ -F ^\r _{\ \a}F^\s _{\
\b}F_{\r\m} F_{\s\n} -A_\a (\pa ^\r F_{\r\s})F_{\m\n}F^\s _{\
\b}\ncr &+& 2A_\m (\pa ^\s A_\n )(\pa _\b A_\m )(\pa ^\r F_{\r\s})
-2A_\a (\pa ^\r F_{\r\s})F_{\b\m}F^\s _{\ \n}+ A_\m (\pa _\n
F^{\r\s})F_{\r\a}F_{\s \b} \ncr &-&A_\m (\pa _\a A_\r)(\pa _\n
F^{\r\s})F_{\s\b} - A_\a (\pa _\b A^\s)(\pa _\m A^{\r})(\pa _\n
F_{\r\s})\ncr
 &+&A_\a A_\m (\pa ^\r F^\s_{\ \b})(\pa _\n F_{\r\s}) +
A_\a F_{\s\b}F_{\m\r}(\pa _\n F^{\r\s}) \Big]\ ,\ncr \label{l2psi}
S_{\psi}^{(2)} &=&\int d^4x\,\Big[ \bar\psi^{(2)}i\g ^\m (\pa
_\m\psi )+\bar\psi i\g ^\m (\pa _\m\psi ^{(2)})+\bar\psi^{(1)}i\g
^\m (\pa _\m\psi ^{(1)})\\ &-&m \bar\psi^{(2)}\psi -m\bar\psi \psi
^{(2)}-m\bar\psi^{(1)}\psi ^{(1)}\ncr &+& \bar\psi^{(2)}\g ^\m A
_\m\psi + \bar\psi \g ^\m A^{(2)} _\m\psi + \bar\psi \g ^\m
 A_\m\psi ^{(2)}+\bar\psi^{(1)}\g ^\m A ^{(1)}_\m\psi + \bar\psi \g ^\m
A^{(1)} _\m\psi ^{(1)}+ \bar\psi ^{(1)}\g ^\m
 A_\m\psi ^{(1)}\ncr
 &+&{i\over 2}\, \theta ^{\a\b}\big[ \bar\psi^{(1)}\g ^\m (\pa _\a A _\m)(\pa _\b\psi )
  + \bar\psi \g ^\m
(\pa _\a A^{(1)} _\m)(\pa _\b\psi )+ \bar\psi \g ^\m (\pa _\a A_\m
)(\pa _\b \psi ^{(1)})\big] \ncr &-&{1\over 8}\,\theta
^{\a\b}\theta^{\r\s}\bar\psi \g ^\m(\pa _\a\pa _\r A_\m )(\pa
_\b\pa _\s \psi ) \Big]\ . \nonumber\eea
 The 'commutative' covariant
derivative is $D_\m \psi =\pa _\m\psi -iA_\m\psi\ $. Notice that
$S_{A}^{(2)}$ is quartic in the gauge potentials, while
$S_{\psi}^{(2)}$ contains terms with 3, 4 and 5 fields. For
example, the part of $S_{\psi}^{(2)}$ with 3 fields is \bea
\label{3psi2} S_{\psi ,3}^{(2)}&=&\int d^4x\, {1\over 8}\, \theta
^{\a\b}\theta ^{\r\s}\Big[ (\pa _\a\pa _\r A_\s)\bar\psi\g ^\m
(\pa _\b \pa _\m\psi )+ (\pa _\a\pa _\m A_\r)\bar\psi\g ^\m (\pa
_\b \pa _\s\psi ) \ncr &-&(\pa _\a\pa _\r A_\m)\bar\psi\g ^\m (\pa
_\b \pa _\s\psi )+ im\, (\pa _\a\pa _\r A_\s)\bar\psi (\pa _\b
\psi )\Big] \ . \eea
 For the purpose of
functional integration, we express the Dirac spinor in terms of
the Majorana spinors. They are introduced as $\psi _{1,2}={1\over
2}(\psi\pm\psi ^C)$, where $\psi ^C=C\bar\psi^T$ is the
charge-conjugated spinor. The Dirac spinor is  $\psi = \psi _1
+i\psi _2\,$; the actions in terms of Majorana spinors are of the
form \bea S ^{(0)} &=& \int d^4x\Big[\bar\psi _1 \big( i\g ^\m \pa
_ \m -m \big)\psi _1 + \bar\psi _2\big( i\g ^\m \pa _\m - m
\big)\psi _2\ncr &+& i\bar\psi _1\g ^\m A_\m\psi _2-i\bar\psi _2\g
^\m A_\m\psi _1-{1\over 4} F^{\m\n}F_{\m\n}\Big] \ ,\\
  S _{\psi
}^{(1)} &=& {1\over 2}\, \theta ^{\r\s} \int d^4x \Big[\big(
-{i\over 2}\bar \psi _1\g ^\m \big( F_{\m\r}\pa _\s + F_{\r\s}\pa
_\m + F_{\s\m}\pa _\r \big) \psi _1 + {1\over 2}\, m F_{\r\s}\bar
\psi _1 \psi _1 \ncr &-& {i\over 2}\, \bar \psi _2\g ^\m \big(
F_{\m\r} \pa _\s + F_{\r\s} \pa _\m + F_{\s\m} \pa _\r \big) \psi
_2 + {1\over 2}\, m F_{\r\s}\bar \psi _2 \psi _2 \ncr &-& {i\over
2}\bar \psi _1\g ^\m \big( F_{\m\r} A_\s + F_{\r\s}A_\m + F_{\s\m}
A_\r \big) \psi _2  \ncr &+&{i\over 2}\bar \psi _2\g ^\m \big(
F_{\m\r} A_\s + F_{\r\s}A_\m + F_{\s\m} A_\r \big) \psi _1 \Big] \
\, ,\ {\rm etc.} \label{llp}\eea The cyclic combinations which
appear in (\ref{llp}) will be  in the following written in a
compact way
 introducing the symbol
$\DG$, e.g.
  $  F_{\m\r}\pa _\s +
F_{\r\s}\pa _\m +F_{\s\m}\pa _\r ={1\over 2}\,\Delta
^{\a\b\g}_{\s\r\m}F_{\g\b}\pa _\a \ .$  $\DG$ is cyclic separately
in upper and lower indices, and antisymmetric in any pair of upper
or of lower indices: \be \Delta ^{\a\b\g}_{\s\r\m} =\d
^\a_\s\d^\b_\r\d^\g_\m -\d ^\a_\r\d^\b_\s\d^\g_\m +({\rm cyclic\
}\a \b \g  )=-\epsilon ^{\a\b\g\l}\epsilon _{\s\r\m\l}\ .\ee The
second order actions can be expressed likewise.

\section{Background field method}

 As explained in \cite{BR}, in order to find the one-loop
effective action via the background field method (and hence the
divergencies at one loop), one expresses all fields $\phi ^i$ as
the sum of the corresponding classical configuration and the
quantum fluctuation: $\phi^i=\phi ^i_0+\Phi^i $. The one-loop
effective action is a functional supertrace: \be \label{G1} \G
[\phi^i_0] =S[\phi^i _0]-{1 \over 2i} {\rm STr} \Big(\log S^{(2)
}[\phi ^i_0]\Big)\ , \ee where $S^{(2)}_{ij}={\d_L \over \d
\phi^i}{\d_R\over \d\phi^j}S\Big|_{\phi ^i=\phi^i_0}$, and ${\d
\over \d \phi^i}$ denotes the functional derivative.

In our case the fields  are the real vector gauge field $A_\m$ and
the Dirac spinor $\psi$, and they are coupled. In order to perform
the functional integration we have to put them into one
'multiplet' field. However,  $A_\m$ is real-number valued while
$\psi$ is complex-Grassmann (if they were independent they
 would have entered  the effective action with different
coefficients $-{1\over 2}$ and 1). To make all fields 'real' we
need to express Dirac spinor $\psi$ in terms of two Majorana
spinors $\psi _1$ and $\psi _2$. Denoting the quantum corrections
by $\ca ^\m$ and $\Psi$ and splitting \be A^\m\to A^\m+\ca^\m\ \
,\ \psi \to \psi +\Psi \ ,\ee we obtain for the quadratic part of
the action
 the expression of the type
\be S^{(2)}=\int d^4x {\pmatrix { \ca_\a & \bar \Psi_1 & \bar
\Psi_2 \cr} }\,\cb\, {\pmatrix{ \ca_\b \cr   \Psi_1 \cr  \Psi_2
\cr}} \ ,\ee where the matrix $\cb$ contains only classical
fields. We have to include in $\cb$ the gauge fixing term  \be
S_{\rm{GF}}=-{1\over 2}\int d^4x(\pa_\m\ca^\m)^2\ ,\ee while the
ghost action will not contribute.
 The
one-loop effective action is  then $$\G_1={i\over 2}\log\Sdet \cb
= {i\over 2}\,\STr\log\cb \ .$$ Matrix $\cb$ can be written as
3$\times$3 block-matrix $$\cb =\pmatrix{ \cb _{11} &\cb _{12} &\cb
_{13} \cr \cb _{21} &\cb _{22} &\cb _{23} \cr \cb _{31} &\cb _{32}
&\cb _{33} \cr } \ ,$$ where the submatrices $\cb _{12}$, $\cb
_{13}$, $\cb _{21}$ and $\cb _{31}$ are Grassmann-odd while the
rest are Grassmann-even. The supertrace of $\cb$ is defined by $
\STr\cb = \Tr\cb _{11}-\Tr\cb _{22} -\Tr\cb _{33}$, and  $\cb$ can
be written as
 $$\cb =\pmatrix{ {1\over 2}g_{\a\b}\Box &0 &0 \cr 0
&i\slash \pa &0 \cr 0&0 &i\slash \pa \cr }+\cm \ .$$
 In order to expand $\log \cb$ around identity,
we multiply it by $\cc\cc^{-1}$ \cite{bv}, with $$\cc =\pmatrix{ 2
& 0 & 0 \cr 0 & -i\slash\pa & 0  \cr 0 & 0 & -i\slash\pa \cr} \
.$$ Then \bea\G_1&=&{i\over 2}\,\STr\log (\cb \cc)+{i\over
2}\,\STr\log \cc^{-1}\\ &=&{i\over 2}\,\STr\log ( \ci +\Box
^{-1}\cm\cc )+ {i\over 2}\,\STr\log \cc^{-1}+{i\over 2}\,\STr\log
\Box  ,\nonumber \eea where $\ci ={\rm diag} (g_{\m\n},1,1)$. As
usual, the second and the third terms, being independent on the
fields, can be included in infinite renormalization. Note that the
propagator for all fields is now $\Box ^{-1}$, while the massive
fermionic terms are in the interaction part, $\cm$.

Performing the transformations described above, for NC QED we
obtain the effective action in the following form: \be \Gamma
=S_0+{i\over 2}\,\STr\log \Big(\ci +\Box ^{-1}N_0 +\Box ^{-1}N_1
+\Box ^{-1}T_1 +\Box ^{-1}T_2+\Box ^{-1}\Pi \Big) \ .\label{str}
\ee Matrices $N_0$, $N_1$, $T_1$ and $T_2$ are given by \be
\label{n0} N_0=\pmatrix {0& 0&0\cr 0&i m \slash\pa & 0\cr 0 &0 & i
m\slash\pa }\ ,\ee

\be\label{n1} N_1=\pmatrix {0& -i\bar\psi\g ^\a \slash\partial &
\bar\psi\g ^\a \slash\partial \cr 2\g ^\b\psi & 0 & \slash
A\slash\pa \cr -2i\g ^\b\psi& -\slash A\slash\pa & 0 }\ ,\ee

$$ T_2=\theta ^{\rho\sigma}\Delta^{\a\b\g}_{\sigma\r\m} \pmatrix
{-{1\over 2}\bar\psi\g ^\m\psi\pa _\g & {i\over 4}({1\over
2}F_{\g\b}+\pa _\g A_\b )\bar\psi\g ^\m \slash\partial &-{1\over
4}({1\over 2}F_{\g\b}+\pa _\g A_\b )\bar\psi\g ^\m \slash\partial
\cr - {1\over 2}\g ^\m\psi (-{1\over 2}F_{\g\a}+A_\a \pa _\g  )
&{i\over 8}A_\a F_{\g\b}\g^\m\slash\pa &-{1\over 8}A_\a
F_{\g\b}\g^\m\slash\pa  \cr {i\over 2}\g ^\m\psi (-{1\over
2}F_{\g\a}+A_\a \pa _\g  ) & {1\over 8}A_\a F_{\g\b}\g^\m\slash\pa
& {i\over 8}A_\a F_{\g\b}\g^\m \slash\pa \cr }\ ,$$

\be\label{t1} T_1=\pmatrix{V& A_1 & A_2\cr
 B_1 &  C & 0 \cr B_2 & 0 & C \cr }\ee
with $A_{1,2},B_{1,2}\ $ and $C$ defined as
 \bea A_{1,2} &=&{i\over 4}\theta
^{\r\s}\Delta ^{\a\b\g}_{\s\r\m}\pa _\g \Big(-(\pa
_\b\bar\psi_{1,2} )i\g ^\m-{m\over 2}\d ^\m _\b
\bar\psi_{1,2}\Big)\slash\pa \ncr
 B_{1,2}
&=&{1\over 2}\theta ^{\r\s}\Delta ^{\a\b\g}_{\s\r\m} \Big( -i\g
^\m (\pa _\a\psi_{1,2})+{m\over 2}\d ^\m _\a \psi_{1,2}\Big)\pa
_\g \ncr
 C
&=&-{i\over 4}\theta ^{\r\s}\Big(- {i\over 2}\Delta
^{\a\b\g}_{\s\r\m}\g ^\m F_{\g\b}\pa _\a +m F_{\r\s}\Big)
\slash\pa \ .\nonumber\eea  $V= \overleftarrow { \pa _\m}
V^{\m\a,\n\b}(x)\overrightarrow { \pa _\b}$ \  comes from the term
$(\pa_\m\ca_\a)V^{\m\a,\n\b}(\pa_\n\ca_\b)$ in $S^{(2)}$, with
\bea \label{v} V^{\m\r,\n\s}&=& {1\over 2}\,
(g^{\m\n}g^{\r\s}-g^{\m\s}g^{\n\r})\theta^{\a\b}F_{\a\b}\ncr
&+&g^{\m\n}(\theta^{\a\r}{F^\s}_\a+\theta^{\a\s}{F^\r}_\a)
+g^{\r\s}(\theta^{\a\m}{F^\n} _\a+\theta^{\a\n}{F^\m} _\a )\ncr
&-&g^{\m\s}(\theta^{\a\r}{F^\n}_\a+\theta^{\a\n}{F^\r}_\a)
-g^{\n\r}(\theta^{\a\s}{F^\m} _\a+\theta^{\a\m}{F^\s} _\a )\ncr
&+& \theta^{\m\r}F^{\n\s}+\theta^{\n\s}F^{\m\r}-
\theta^{\r\s}F^{\m\n} -\theta^{\m\n}F^{\r\s}
-\theta^{\n\r}F^{\m\s} -\theta^{\m\s}F^{\n\r} \ .\nonumber\eea The
terms $T_1$ and $T_2$ are linear in $\theta$. $\Pi $ denotes the
contribution which comes from the $\theta ^2$-classical
lagrangian. It can be split as $\Pi =\Pi _1+\Pi _2 +\Pi _3$; as
for $T$'s, index denotes the number of background fields  present
in the corresponding matrix. E.g. from (\ref{3psi2}) we get for
$\Pi _1$ : \be\label{p1} \Pi _1=\pmatrix{0& -iR_2\slash\pa &
-iR_1\slash\pa \cr
 2S_2 &  0 &  -iP\slash\pa \cr  -2S_1 & iP\slash\pa & 0 \cr }\ee
with \bea R_{1,2}&=&{1\over 8}\,\theta ^{\m\n}\theta ^{\r\s}\Big[
-i\d ^\a _\s\pa _ \m\pa _\r\bar\psi _{1,2}\slash\pa\pa _\n-i\d ^\a
_ \r\pa _\m\pa _\b \bar\psi _{1,2}\g ^\b\pa _\s\pa _\n \ncr &+&i\d
^\a _{\b}\pa _\m\pa _\r \bar\psi _{1,2}\g ^\b\pa _\s\pa _\n +m\d
^a _{\s}\pa _\m\pa _\r\bar\psi _{1,2}\pa _\n\Big]\ ,\ncr
 S_{1,2}&=&{1\over 8}\,\theta ^{\m\n}\theta ^{\r\s}\Big[
i\d ^\a_{\s}(\pa _ \n\slash \pa \psi _{1,2})\pa _\m\pa _\r +i\d
^\a _{\r}\g ^\b ( \pa _\n\pa _\s\psi _{1,2})\pa _\m\pa _\b \ncr
&-&i\d ^\a_{\b}\g ^\b (\pa _\n\pa _\s \psi _{1,2})\pa _\m\pa _\r
-m\d ^\a_{\s}(\pa _\n \psi _{1,2})\pa _\m\pa _\r \Big] \ ,\ncr
 P&=&{1\over 8}\,\theta ^{\m\n}\theta ^{\r\s}\Big[ {1\over 2}\,(\pa _\m
 F_{\r\s})(i\slash\pa -m)\pa _\n -i(\pa _\m F_{\r\b })\g ^\b\pa
 _\n\pa _\s
\Big]\ .\eea

\section{Divergent one-loop effective action}

The operator $\cb\cc$  in the formula (\ref{str}) is split in a
way convenient for the analysis of  perturbation series. Let us
explain the notation again.  $T$-matrices are linear in the
parameter $\theta$, $\Pi$ are quadratic in $\theta$. Index
 denotes the number of classical fields in a given matrix,
i.e. in diagrammatic language, shows the number of 'external
legs' of the corresponding diagram. In our calculation we confine
to the corrections of linear and quadratic
 order in $\theta$ and of the second order in
classical fields. If we consider the expansion of (\ref{str}) \bea
\label{strexp} \G_1&=&{i\over 2}\,\STr\log \big( \ci +\Box
^{-1}N_0 +\Box ^{-1}N_1 +\Box ^{-1}T_1 +\Box ^{-1}T_2 +\Box
^{-1}\Pi _1 +\Box ^{-1}\Pi _2 +\Box ^{-1}\Pi _3 \big)\ncr
&=&{i\over 2} \sum _{n=1}^\infty {(-1)^{n+1}\over n}\,
\STr\big(\Box ^{-1}N_0 +\Box ^{-1}N_1 +\Box ^{-1}T_1 +\Box
^{-1}T_2 +\Box ^{-1}\Pi\big) ^n \ ,\eea it seems as we have to
include only the powers $n=1,2$. But due to the nonvanishing
fermionic mass $m$ (i.e. the existence of the term $N_0$), in
principle we will have to take into account also higher powers of
$n$. $n$ is finite and determined by the fact that we want to
calculate only the divergent part. Analyzing the structure of $\nn
,\dots ,\td $ in some detail we conclude that the following terms
(from $\theta$-linear action) in the expansion (\ref{strexp}) may
be divergent: $(\nn )^k\td  $ for $k=2,3,4$; $(\nn )^k\nj\tj$ for
$k=1,2,3$;
 $(\nn )^k(\nj )^2$ for $k=1,2$ and
 $(\nn )^k(\tj )^2$ for $k=1,2,3,4$.
(Here, of course, terms are written symbolically, i.e. without the
exact order of the operators.) From $\theta$-quadratic action the
contribution comes from $\Box ^{-1}N_1 \Box ^{-1}\Pi _1$ and
 $\Box ^{-1}N_0 \Box ^{-1}\Pi _2$; $\Pi _3$ contributes
 only to vertices.
 It is also clear that in the massless fermionic
 case the absence of $N_0$ brings a considerable simplification.

In order to compare with the known result \cite{bgpsw1}, we first
calculate the divergencies in the purely bosonic case: we put
 $N_0=0$, $N_1=0$, $T_2=0$ and $\Pi _1=0$. The contribution coming
 from $\Pi _2$ vanishes in all cases. Assuming all this,
$T_1$ reduces to $\ct _1$: \be \ct _1 =\pmatrix{ V &0 &0\cr
0&0&0\cr 0&0&0\cr}\ .\ee Denoting \be \G_b={i\over 2}\,\STr\log
\big( 1+\Box ^{-1}\ct _1\big) = {i\over 2}\Big[\Tr \Box ^{-1}\ct
_1 -{1\over 2}\Tr (\Box ^{-1}\ct _1 )^2+\dots\Big] \label{vv}\ee
we get, after dimensional regularization of the traces, extraction
of the divergent parts and multiple use of partial integration and
Bianchi identities, \be \G_{b}={1\over 64\pi^2\e}\,\int
d^4x\Big[\Box \tilde F^{\m\n}\Box\tilde F_{\m\n}+{2\over
15}\,\Box\tilde F^{\m\r}\pa_\m\pa^\n\tilde F_{\n\r}+{1\over
5}\,\Box F^{\m\n}\tilde \pa_a\tilde \pa^\a F_{\m\n}-{1\over
4}\,\theta ^2\Box F^{\m\n}\Box F_{\m\n}\Big]\
.\label{edkrajnje}\ee We use the notation $\tilde F^{\m\n}={\theta
^\m}_\a F^{\a\n}$, $\ \tilde F =\theta _{\m\n}F^{\m\n}$, $\
\tilde\pa ^\m ={\theta ^\m}_\a \pa ^\a$, $\ \theta ^2 =\theta
^{\r\s}\theta _ {\r\s}$.

For the full NC QED case, from the supertraces containing $T_1$
and $T_2$, we obtain \bea\G_1^{(1)}&=&{1\over (4\pi)^2\e}\int d^4
x \Big[ 4i\bar\psi\slash\pa\psi-16m\bar\psi\psi-{2\over
3}F_{\m\n}F^{\m\n}\\ \nonumber &+& \theta^{\a\b}\Big( {1\over
3}\,\bar\psi\g _\a\pa _\b (\Box -im\slash\pa )\psi +{1\over
6}\bar\psi\s _{\a\b}\Box (i\slash\pa -m)\psi\\ \nonumber
&+&m^2\bar\psi\g_\a\pa_\b\psi+{m^2\over 2}\,\bar
\psi\s_{\a\b}(i\slash\pa-m)\psi \Big)\\ \nonumber &-&{1\over
120}\,\tilde F^{\r\s}\Box ^2\tilde F_{\s\r}+{1\over 30}\,\tilde
F^{\r\s}\Box ^2\tilde F_{\r\s} -{1\over 30}\,\tilde
F^{\r\s}\Box\pa _\s\pa^\n\tilde F_{\r\n}\\ \nonumber
 &+&{m^2\over
6}\,\tilde F^{\r\s}\Box \tilde F_{\r\s}-{m^2\over 12}\,\tilde
F^{\r\s}\Box \tilde F_{\s\r} -{m^2\over 6}\,\tilde F^{\r\m}\pa
_\m\pa^\n\tilde F_{\r\n} -{m^4\over 4}\,\tilde F_{\r\s}\tilde
F^{\s\r}\\ \nonumber &+&{i\over
48}\,\theta^2\bar\psi\Box^2\slash\pa\psi-{i\over
24}\,\theta^{\a\m}\theta^\b_{\
\m}\bar\psi\Box\slash\pa\pa_\a\pa_\b\psi -{i\over
12}\,\theta^{\a\m}\theta^\b_{\ \m}\bar\psi\Box^2\g_\a\pa_\b\psi \\
\nonumber &+&{m\over 12}\,\theta^{\a\m}\theta ^b_{\
\m}\bar\psi\pa_\r\pa_\s\Box\psi +{5im^2\over
48}\,\theta^2\bar\psi\Box\slash\pa\psi-{im^2\over 24}\,
\theta^{\a\m}\theta^\b_{\ \m}\bar\psi\slash\pa\pa_\a\pa_\b\psi
-{7im^2\over 24}\,\theta^{\a\m}\theta^\b_{\
\m}\bar\psi\g_\a\pa_\b\Box\psi\\ \nonumber &+&{m^3\over
4}\,\theta^{\a\m}\theta^\b_{\ \m}\bar\psi\pa_\a\pa_\b\psi
+{5im^4\over 24}\,\theta^2\bar\psi\slash\pa\psi-{im^4\over
3}\theta^{\a\m}\theta^\b_{\ \m}\bar\psi\g_\a\pa_\b\psi -{m^5\over
8}\,\theta ^2\bar\psi\psi \Big]+\G_{b}\ .\label{krajnje}\eea The
traces containing $\Pi $'s give additionally
 \bea \G_1^{(2)}&=& {1\over 4(4\pi)^2\e}\int d^4 x \Big[
{1\over 240}\tilde F\Box ^2\tilde F+ {1\over 120}\,\tilde
F^{\r\s}\Box ^2\tilde F_{\s\r}-{1\over 15}\,\tilde F^{\r\s}\Box
^2\tilde F_{\r\s}\\ \nonumber  &+&{1\over 15}\,\tilde
F^{\r\s}\Box\pa _\s\pa^\n\tilde F_{\r\n}+{im\over 24}\theta
^{\a\m}\theta _\a^{\ \n}\bar\psi\,\sigma_{\b\n}\pa ^\b\pa
_\m\Box\psi \Big] \ . \eea

\section{Discussion}

Our goal in this lecture was to obtain the divergent part of the
one-loop effective action  in NC QED in the second order in the
noncommutativity parameter $\theta$ and the same order in the
classical fields, $\psi ,\ A^\m$. Thus we obtained the second
order corrections to the propagators in the theory and therefore
the form of the counterterms necessary for renormalization. The
method we used is the background field method; the initial point
for the perturbative expansion is (\ref{str}). Expansion is
written in such a way that it is easy to sample out the terms
contributing to the 2-point, 3-point, 4-point etc. functions.

The main motive of this calculation was to check
 the renormalizability of $\theta$-expanded NC QED in the first
and second order in $\theta$, and the possibilities of
 generalization  to all orders. This
was done for the pure NC U(1) in \cite{bggpsw}. The trick which
was used is that the SW map is not unique and thus does not fix
the fields in the $\theta$-expansion fully, but allows for their
redefinitions. If the fields are expanded (written symbolically)
as \be \label{SWexpansion}\hat A_\m=\sum \theta ^n A_\m^{(n)}\ ,\
\ \hat \psi =\sum \theta ^n \psi ^{(n)}\ ,\ee the allowed
redefinitions can be of the form \be\label{reda}
{A_\m^{(n)}}^\prime =A_\m^{(n)} + {\bf A}_\m^{(n)}\ ,\ee
\be\label{redpsi} {\psi ^{(n)}}^\prime =\psi ^{(n)} + {\bf
\Psi}^{(n)}\ ,\ee where ${\bf A}_\m^{(n)}$, ${\bf \Psi} ^{(n)}$
are gauge covariant expressions of appropriate dimension with
exactly $n$ factors of $\theta$. These field redefinitions produce
in the actions the
 extra-terms of the following forms \cite{bggpsw}:
\bea \label{DSA} \Delta S_A &=&\int d^4x\, (D_\n F^{\m\n}){\bf
A}_\n^{(n)}\\ \label{DSpsi} \Delta S_\psi &=&\int d^4x\, \Big[
\bar\psi (i\slash D -m){\bf \Psi}^{(n)}+ \bar {\bf
\Psi}^{(n)}(i\slash D -m)\psi \Big]\ .\eea So, if the
renormalizability of the theory can be achieved by the field
redefinitions, all counterterms have to be of the types
(\ref{DSA}-\ref{DSpsi}), and in the final theory we will have only
the redefined, 'physical' fields and no divergencies.

It is easy to see that in the purely bosonic case the action
(\ref{edkrajnje}) is of the type (\ref{DSA}). The gauge potential
can be redefined into physical potential: $A_\a \to A_\a +{\bf
A}_\a^{(2)}$: \be{\bf A}_\a^{(2)}=\jcpe\Big[- {1\over 120}\Box
^2\tilde \pa ^\r\tilde F^{\a\r} +{17\over 60}\Box ^2\tilde \pa
^\r\tilde F^{\r\a} -{1\over 20}\Box\pa ^\r\tilde\pa ^\m\tilde\pa
_\m F_{\r\a}+{1\over 16}\theta ^2\Box ^2\pa ^\r F_{\r\a}\Big] \ ,
\ee and the divergent term in the effective action will cancel
with the one coming from the field redefinition.

 Let us discuss what
happens when the fermions are present. All bosonic corrections
(which come from fermionic parts in the trace) are of the $\theta
^2$-order, and all are of  allowed type except for the term
$-\jcpe {m^4\over 4}\tilde F_{\r\s}\tilde F^{\s\r}$ . The
fermionic $\theta$-linear correction, on the other hand, is \be
\label{FC}\jcpe \theta^{\a\b}\Big[ {1\over 3}\,\bar\psi\g _\a\pa
_\b (\Box -im\slash\pa )\psi +{1\over 6}\bar\psi\s _{\a\b}\Box
(i\slash\pa -m)\psi +m^2\bar\psi\g_\a\pa_\b\psi+{m^2\over 2}\,\bar
\psi\s_{\a\b}(i\slash\pa-m)\psi \Big]\ .\ee One can check that the
pieces \ $\, {1\over 3}\,\bar\psi\g _\a\pa _\b \Box \psi +{1\over
6}\,\bar\psi\s _{\a\b}\Box (i\slash\pa -m)\psi\, $\ and \ $\,
m^2\bar\psi\g_\a\pa_\b\psi+{m^2\over 2}\,\bar
\psi\s_{\a\b}(i\slash\pa-m)\psi\, $\  of (\ref{FC}) can be
obtained from the field redefinitions, leaving the term\
$-{im\over 3}\bar\psi\gamma _\r\pa _\s\psi$ \ excessive. Thus we
see that renormalizability cannot be achieved in the massive case
$m\neq 0$, as there are (at least) two terms which obstruct it.

One observes further that for $m=0$  the $\theta ^2$ fermionic
contribution fits into the field redefinition scheme, too. This
contribution reads \be \jcpe\Big[ {i\over
48}\,\theta^2\bar\psi\Box^2\slash\pa\psi-{i\over
24}\,\theta^{\a\m}\theta^\b_{\
\m}\bar\psi\Box\slash\pa\pa_\a\pa_\b\psi -{i\over
12}\,\theta^{\a\m}\theta^\b_{\ \m}\bar\psi\Box^2\g_\a\pa_\b\psi
\Big]\ ,\ee and it is of the form (\ref{DSpsi}). The full field
redefinition of fermions in the massless case is  given by \be
{\bf \Psi}^{(1)}= \jcpe\theta^{\a\b}\Big[ {1\over 6}\sigma
_{\m\a}\pa _\b\pa ^\m\psi +{1\over 12}\sigma _{\a\b}\Box\psi
\Big]\ ,\ee
\be
{\bf \Psi}^{(2)}= \jcpe\Big[ {1\over 96}\,\theta^2
\Box^2\psi-{1\over 16}\,\theta^{\a\m}\theta^\b_{\
\m}\Box\pa_\a\pa_\b\psi +{i\over 24}\,\theta^{\a\m}\theta^\b_{\
\m}\sigma _{\m\a}\Box\pa ^\m\a\pa_\b\psi \Big]\ \ee and it renders
the propagators of physical fermions finite to $\theta ^2$ order.
It is easy to extend this conclusion to higher orders. Namely, the
higher-order divergencies for the propagators ($m=0$) after the
partial integration have the form \bea \label{under}
S_A^{(n)}&=&\underbrace{\theta\dots\theta}_n\,A\,\underbrace{\pa\dots\pa}_k
A\ncr S_\psi
^{(n)}&=&\underbrace{\theta\dots\theta}_n\,\bar\psi\,\g\dots\,
\underbrace{\pa\dots\pa}_l \psi \ .\eea Counting dimensions, we
see that $k=2+2n$, $l=1+2n$, so there is always at least one
derivative in the expressions (\ref{under}). This in the massless
case (along with the fact that everything, being a result of the
background field method, is covariant) ensures that they can be
interpreted as field redefinitions.

Unfortunately, as also stressed in \cite{w}, the full
renormalizability is spoiled by vertex corrections. Obviously,
$\theta$-expansion introduces new interactions into NC QED
(compared to the commutative QED), like photon self-interaction
and four-fermion interaction. The divergent part of the 4-fermion
vertex, in the linear order in $\theta$, is found to be
\be\label{4psi} S_{4\psi} =-\jcpe \,{1\over 2}\,\theta
^{\r\s}\DG\int d^4x\, \bar\psi\g ^\m\psi\, \bar\psi \g _ \a\g
_\b\g _\g\psi =-\jcpe\, 3i\theta ^{\r\s}\int d^4x\epsilon
_{\r\s\m\n}  \bar\psi\g ^\m\psi\, \bar\psi \g _5\g ^\n\psi \ .\ee
This term is independent of $m$ and cannot be achieved by field
redefinition. Thus we see that it spoils renormalizability  in the
massless fermionic case. Finally, even if all vertex divergencies
were of the prescribed form, one would have to check (a quite
nontrivial property)  that redefinitions would remove the
divergencies in all propagators and vertices simultaneously.

So we are left with the conclusion that NC QED is not
renormalizable and can only be treated as an effective theory at
low energies. The regularization procedure than gives new
effective interactions, which have to be included in the cross
sections. Many interesting phenomenological questions arise: e.g.
can $4\psi$ vertex  be interpreted in terms of some effective
(scalar) particles,  higgses? In order to investigate this
possibility in more details, it would be useful to calculate the
$4\psi$ contribution in the $\theta ^2$ order for $U(1)$, or in
$\theta$-linear order for some other gauge theory like $SU(2)$.

\end{document}